\documentclass[letter,twocolumn]{jpsj3}
\usepackage{txfonts}
\usepackage{color}
\def\gsim{\mathop {\vtop {\ialign {##\crcr 
$\hfil \displaystyle {>}\hfil $\crcr \noalign {\kern1pt \nointerlineskip } 
$\,\sim$ \crcr \noalign {\kern1pt}}}}\limits}
\def\lsim{\mathop {\vtop {\ialign {##\crcr 
$\hfil \displaystyle {<}\hfil $\crcr \noalign {\kern1pt \nointerlineskip } 
$\,\,\sim$ \crcr \noalign {\kern1pt}}}}\limits}

\title{Origin of Quantum Criticality in Yb-Al-Au 
Approximant Crystal and Quasicrystal
}

\author{Shinji Watanabe$^1$ and Kazumasa Miyake$^2$}
\inst{$^1$Department of Basic Sciences, Kyushu Institute of Technology, Kitakyushu, Fukuoka 804-8550, Japan \\
$^2$Toyota Physical and Chemical Research Institute, Nagakute, Aichi 480-1192, Japan 
} 

\abst{
To get insight into the mechanism of emergence of unconventional quantum criticality observed in quasicrystal Yb$_{15}$Al$_{34}$Au$_{51}$, the approximant crystal Yb$_{14}$Al$_{35}$Au$_{51}$ is analyzed theoretically.  
By constructing a minimal model for the approximant crystal, 
the heavy quasiparticle band is shown to emerge near the Fermi level because of 
strong correlation of 4f electrons at Yb. 
We find that charge-transfer mode between 4f electron at Yb on the 3rd shell and 3p electron 
at Al on the 4th shell in Tsai-type cluster is considerably enhanced with almost flat momentum dependence. 
The mode-coupling theory shows that magnetic as well as valence susceptibility exhibits $\chi\sim T^{-0.5}$ 
for zero-field limit and is expressed as a single scaling function of the ratio of temperature to magnetic field $T/B$ 
over four decades even in the approximant crystal when some condition is satisfied by varying parameters, e.g., by applying pressure. 
The key origin is clarified to be due to 
the strong locality of the critical Yb-valence fluctuation and small Brillouin zone 
reflecting the large unit cell, giving rise to the extremely-small characteristic energy scale. 
This also gives a natural explanation for the quantum criticality in the quasicrystal corresponding to the infinite limit of the unit-cell size. 
}

\begin{document}
\maketitle

Recent discovery of unconventional quantum critical phenomena in quasicrystal Yb$_{15}$Al$_{34}$Au$_{51}$ has attracted much attention~\cite{Deguchi,Watanuki}. 
The measured criticality such as magnetic susceptibility 
$\chi\sim T^{-0.5}$, NMR relaxation rate $(T_{1}T)^{-1}\sim T^{-0.5}$, 
specific heat $C_{\rm e}/T\sim-\log{T}$, and resistivity $\rho\sim T$ 
is unconventional and quite similar to those observed in periodic crystals 
such as heavy-electron metals 
YbRh$_2$Si$_2$~\cite{Gegenwart} and $\beta$-YbAlB$_4$~\cite{Nakatsuji,Matsumoto}, 
which are well explained by quantum critical phenomena of Yb-valence fluctuations~\cite{WM2010}. 

\begin{figure}
\includegraphics[width=7.5cm]{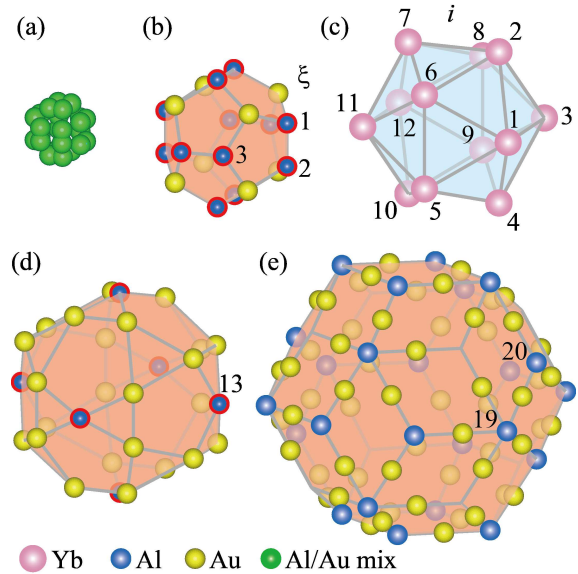}
\caption{(color online) Concentric shell structures of Tsai-type cluster in the Yb-Al-Au approximant:
(a) first shell, (b) second shell, (c) third shell, (d) fourth shell, and (e) fifth shell. 
The number in (c) indicates the $i$-th Yb site and number 
in (b), (d), and (e) indicates the $\xi$-th Al site.
}
\label{fig:Yb_cluster}
\end{figure}

The quasicrystal Yb$_{15}$Al$_{34}$Au$_{51}$ is constituted of 
concentric shell structures of Tsai-type cluster 
shown in Fig.~\ref{fig:Yb_cluster}~\cite{Ishimasa,Deguchi}. 
There also exists approximant crystal Yb$_{14}$Al$_{35}$Au$_{51}$~\cite{Ishimasa}. 
The approximant crystal has periodic arrangement of the body-centered cubic (bcc) structure 
whose unit cell contains shell structures shown 
in Figs.~\ref{fig:Yb_cluster}(a)-\ref{fig:Yb_cluster}(e). 
Theoretical analysis of the Yb-Al-Au cluster 
taking account of critical Yb-valence fluctuations has provided 
a natural explanation for 
robustness of 
quantum criticality in the quasicrystal measured under pressure 
and 
has pointed out a possibility 
that 
the same criticality appears even in the approximant crystal when pressure is applied~\cite{WM2013,WM2015}. 

On the other hand, a new theoretical framework 
for critical Yb-valence fluctuation under magnetic field 
has been developed recently~\cite{WM2014}. 
The theory has succeeded in explaining novel scaling discovered in $\beta$-YbAlB$_4$
where the magnetic susceptibility $\chi$ is expressed as a single scaling function of the ratio of temperature to magnetic field $T/B$ over four decades~\cite{Matsumoto}. 

Surprisingly, recent measurement of magnetic susceptibility 
in the quasicrystal Yb$_{15}$Al$_{34}$Au$_{51}$ has revealed that 
the same $T/B$-scaling behavior appears over six decades of $T/B$~\cite{Deguchi_TB}.
This strongly suggests that the origin of both the unconventional criticality in  
the quasicrystal and periodic crystal $\beta$-YbAlB$_4$ is the same 
and calls for theoretical analysis from the viewpoint of the critical Yb-valence fluctuation. 

Thus the aim of this Letter is to get insight into the origin of the unconventional criticality 
and the mechanism of emergence of the $T/B$ scaling in the magnetic susceptibility  
observed in the quasicrystal from the viewpoint 
by performing the explicit calculation.
Since the locality of the critical Yb-valence fluctuation is considered to be the key origin, 
essentially the same phenomena are expected to occur in the approximant crystal 
when pressure is tuned~\cite{Matsukawa2016}. 
Hence, we clarify the origin of the unconventional criticality and mechanism in the approximant crystal Yb$_{14}$Al$_{35}$Au$_{51}$.

Let us start with setting the model Hamiltonian. 
Recent measurement performed 
by replacing Al with Ga in the Yb-Al-Au quasicrystal 
has revealed that the quantum critical state disappears~\cite{Matsukawa}.
This suggests that the conduction electron at Al, presumably 3p electron, 
contributes to the quantum critical state. 
Hence, we consider the simplest minimal model for the 4f-hole orbital at the Yb site 
and the conduction-hole orbital at the Al site in the approximant crystal: 
\begin{eqnarray}
H=H_{\rm f}+H_{\rm c}+H_{\rm hyb}. 
\label{eq:PAM}
\end{eqnarray}
The 4f-hole part is given by 
\begin{eqnarray}
H_{\rm f}=
\sum_{j=1}^{N_{\rm L}}
\left[
\varepsilon_{\rm f}\sum_{i=1\sigma}^{24}n_{ji\sigma}^{\rm f}
+U\sum_{i=1}^{24}n_{ji\uparrow}^{\rm f}n_{ji\downarrow}^{\rm f}
\right],
\label{eq:Hf}
\end{eqnarray}
where $f_{ji\sigma}^{\dagger}$ $(f_{ji\sigma})$ is the creation (anihilation) operator 
for f hole  
at the $i$-th site in the $j$-th unit cell with spin $\sigma$, 
and $n_{ji\sigma}^{\rm f}{\equiv}f_{ji\sigma}^{\dagger}f_{ji\sigma}$.  
Here $N_{\rm L}$ is the number of unit cells and $i$ specifies the Yb site  
on the Yb$_{12}$ cluster [see Fig.~\ref{fig:Yb_cluster}(c)] at the body center $(i=1\sim 12)$ 
and at the 8 corners of the bcc unit cell $(i=13\sim 24)$. 
The first term represents the f level and the 2nd term represents the on-site Coulomb repulsion
of the 4f holes at the Yb sites. 

The conduction-hole part is given by 
\begin{eqnarray}
H_{\rm c}=-
\sum_{\langle j{\xi}, j'{\nu}\rangle\sigma}
\left(t_{j{\xi}, j'{\nu}}
c_{j\xi\sigma}^{\dagger}c_{j'\nu\sigma}+{\rm h.c.}
\right), 
\label{eq:Hc}
\end{eqnarray}
where $c_{j\xi\sigma}^{\dagger}$ $(c_{j\xi\sigma})$ is the creation (annihilation) operator 
for conduction hole  
at the $\xi$-th site in the $j$-th unit cell with spin $\sigma$. 
Note that $\xi$ specifies Al sites on the 2nd shell [see Fig.~\ref{fig:Yb_cluster}(b)] and 
the 4th shell [see Fig.~\ref{fig:Yb_cluster}(d)] at the body center and 8 corners 
at the unit cell ($12\times 2$ and $6\times 2$ sites, respectively) and 
on the 5th shell [see Fig.~\ref{fig:Yb_cluster}(e)]  ($12$ sites).  
Here $\langle j{\xi}, j'{\nu}\rangle$ denotes the pairs between the $j\xi$-th Al site and the $j'\nu$-th Al site. 
The transfer integrals $t_{j{\xi}, j'{\nu}}$ are set 
for the nearest-neighbor (N.N.) Al sites on the 2nd shell as $t_{2}$ and 
between the 2nd and 4th shells as $t_{2-4}$, and 
are set up to the next N.N. Al sites 
between the 4th and 5th shells as $t_{4-5}$. 
Since existence ratio of the 1st shell [see Fig.~\ref{fig:Yb_cluster}(a)] 
is quite small (Al/Au:$7.8\%$/$8.9\%$~\cite{Ishimasa}),  
we consider the 2nd-5th shells. 
As a first step of analysis, here we consider the case that 
the Al/Au mixed sites framed in red in Fig.~\ref{fig:Yb_cluster} are occupied by Al 
and degeneracy of the 3p orbital is neglected at Al sites. 
In reality, there may also exist conduction orbital at the Au site. 
To take into account this effect, we consider the effective transfer via the Au site 
as a parameter, which is expressed as 
$t'_{5}$ between the N.N. Al sites on the 5th shell 
since existence ratio of the Al's is $100\%$~\cite{Ishimasa}. 

The hybridization between 4f-hole and conduction-hole orbitals is given by 
\begin{eqnarray}
H_{\rm hyb}=
\sum_{\langle ji, j'\xi\rangle\sigma}
\left(V_{ji, j'\xi}
f_{ji\sigma}^{\dagger}c_{j'\xi\sigma}+{\rm h.c.}
\right), 
\label{eq:Hhyb}
\end{eqnarray}
where $\langle ji, j'\xi\rangle$ denotes the pairs between the $ji$-th site and 
the $j'\xi$-th site  
and the hybridization matrix element is given by $V_{ji, j'\xi}$. 
Here $i$ specifies the Yb site $(i=1\sim 24)$ in the $j$-th unit cell and $\xi$ 
specifies the N.N. Al sites on the 4th and 5th shells 
and up to the next N.N. Al sites on the 2nd shell
in the $j'$-th unit cell for each $i$-th site. 
Corresponding $V_{ji, j'\xi}$ is expressed as $V_{3-4}$, $V_{3-5}$, and $V_{2-3}$, respectively. 

To set transfer integrals, we employ the relation $t_{j\xi, j'\nu}\propto 1/r^{\ell+\ell'+1}$, where 
$r$ is the distance between the $j\xi$-th site and $j'\nu$-th site with wave functions 
with  azimuthal quantum numbers $\ell$ and $\ell'$, respectively~\cite{Andersen1,Andersen2,Harrison}. 
By inputting $\ell=\ell'=1$ and $r$ obtained from Ref.~\citen{Ishimasa}, we have 
$t_{2-4}=0.357t_2$ and $t_{4-5}$ is set to be either of $0.173t_2$ or $0.052t_2$  corresponding to Al-Al distances (for instance, the former is set between the $\xi=13$th and $19$th Al sites and the latter is set between the $\xi=13$th and $20$th Al sites 
in Figs.~\ref{fig:Yb_cluster}(d) and \ref{fig:Yb_cluster}(e), respectively).  
As for the f-c hybridization, the similar relation $V_{ji, j'\xi}\propto 1/r^{\ell+\ell'+1}$ holds. 
Hence by inputting $\ell=3$ and $\ell'=1$ and the Yb-Al distance $r$, 
$V_{3-4}$ and $V_{3-5}$ are set as 
$V_{3-4}=1.070V_0$ and $V_{3-5}=0.714V_0$, respectively.  
$V_{2-3}$ is set to be either of $V_0$ or $0.767V_0$ corresponding to Yb-Al distances 
(for instance, the former is set between the $i=1$st and $\xi=1$st sites and 
the latter is set between the 
$i=1$st and $\xi=3$rd sites in Figs.~\ref{fig:Yb_cluster}(c) and \ref{fig:Yb_cluster}(b), respectively). 
Hereafter, the energy unit is taken as $t_2$, i.e., $t_2=1.0$, 
and  $V_0$ and $t_{5}'$ are set as parameters.  

\begin{figure}
\includegraphics[width=7.5cm]{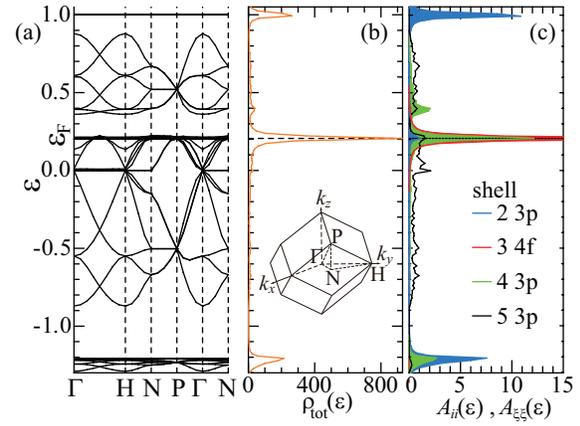}
\caption{(color online) 
(a) Energy band along high-symmetry lines for $t_2=1.0$, $t_{5}'=0.2$, $V_0=0.13$, $\varepsilon_{\rm f}=-0.4$, and $U=\infty$ at $\bar{n}=1$ calculated in 
$N_{\rm L}=8^3$. 
(b) Total density of states. 
Inset shows bcc Brillouin zone. 
(c) Spectral functions 
$A_{ii}(\varepsilon)$ for $i=1$ (red) and 
$A_{\xi\xi}(\varepsilon)$ 
for $\xi=1$ (blue), $\xi=13$ (green), and $\xi=19$ (black). Horizontal dashed lines 
indicate Fermi level $\varepsilon_{\rm F}$. 
}
\label{fig:band_DOS}
\end{figure}

Heavy electron behavior observed in the approximant crystal Yb$_{14}$Al$_{35}$Au$_{51}$~\cite{Deguchi} is considered to be originated from strong onsite Coulomb repulsion $U$ between the 4f holes at Yb.  
To clarify the band structure for the heavy quasiparticles in the approximant crystal, 
we apply the slave-boson mean-field theory~\cite{Read,OM2000} for $U=\infty$ to Eq.~(\ref{eq:PAM}). 
To describe the state for $U=\infty$, we consider $V_{ji, j'\xi}f_{ji\sigma}^{\dagger}b_{i}c_{j'\xi\sigma}$ instead of $V_{ji, j'\xi}f_{ji\sigma}^{\dagger}c_{j'\xi\sigma}$ 
by introducing the slave-boson operator $b_i$ at the $i$-th site 
in the $j$-th unit cell to describe $f^0$ state 
and require the constraint $\sum_{\sigma}n^{\rm f}_{ji\sigma}+b^{\dagger}_{i}b_{i}=1$ with introducing the Lagrange multiplier $\lambda_{i}$, i.e., $\sum_{i}^{24}\lambda_{i}(\sum_{\sigma}n^{\rm f}_{ji\sigma}+b^{\dagger}_{i}b_{i}-1)$. 
We employ the mean-field treatment as $\overline{b_i}=\langle b_{i}\rangle$ 
and the resultant Hamiltonian is denoted by $\tilde{H}$. 
By optimizing the ground-state energy with respect to $\lambda_{i}$ and $\overline{b_i}$, 
$\partial\langle \tilde{H}\rangle/\partial\lambda_{i}=0$ and $\partial\langle \tilde{H}\rangle/\partial\overline{b_{i}}=0$, the following mean-field equations are obtained: 
\begin{eqnarray}
\frac{1}{N_{\rm L}}\sum_{{\bf k}\sigma}
\langle f_{{\bf k}i\sigma}^{\dagger}f_{{\bf k}i\sigma}\rangle
+\overline{b_{i}}^2&=&1, 
\label{eq:MF1}
\\
\frac{1}{2N_{\rm L}}\sum_{{\bf k}\sigma}
\sum_{\xi}
\left[
V_{{\bf k}i\xi}
\langle f^{\dagger}_{{\bf k}i\sigma}c_{{\bf k}\xi\sigma}
\rangle
+{\rm h.c.}
\right]
+\lambda_{i}\overline{b_{i}}&=&0, 
\label{eq:MF2}
\end{eqnarray}
for $i=1, .., 24$.

The filling is defined by the hole number per site, which is given by 
$\bar{n}=\bar{n}_{\rm f}+\bar{n}_{\rm c}$ where $\bar{n}_{\rm f}\equiv\frac{1}{N_{\rm L}}\sum_{j=1}^{N_{\rm L}}\frac{1}{24}\sum_{i=1}^{24}\sum_{\sigma}\langle n_{ji\sigma}^{\rm f}\rangle$ 
and $\bar{n}_{\rm c}\equiv\frac{1}{N_{\rm L}}\sum_{j=1}^{N_{\rm L}}
\frac{1}{48}\sum_{\xi=1}^{48}\sum_{\sigma}\langle n_{j\xi\sigma}^{\rm c}\rangle$ 
with $n_{j\xi\sigma}^{\rm c}\equiv c_{j\xi\sigma}^{\dagger}c_{j\xi\sigma}$. 
We have solved Eqs.~(\ref{eq:MF1}) and (\ref{eq:MF2}), and $\bar{n}=1$ self-consistently at the ground state for several $V_{0}$, $\varepsilon_{\rm f}$, and $t_{5}'$.  
Figure~\ref{fig:band_DOS} shows 
the result for $V_0=0.13$, $\varepsilon_{\rm f}=-0.4$ and $t_{5}'=0.2$ 
calculated in $N_{\rm L}=8\times8\times8$   
as a typical case for the approximant crystal. 

The renormalized f level is raised up to $\varepsilon\approx0.2$, where the Fermi level $\varepsilon_{\rm F}$ is located, giving rise to the heavy quasiparticle band in Fig.~\ref{fig:band_DOS}(a). This is reflected in the sharp peak of the total density of states (DOS) $\rho_{\rm tot}(\varepsilon)$ around $\varepsilon_{\rm F}$, so-called Kondo peak, in Fig.~\ref{fig:band_DOS}(b).  
Here,  
$\rho_{\rm tot}(\varepsilon)$ is given by  
$\rho_{\rm tot}(\varepsilon)=\rho_{\rm f}(\varepsilon)+\rho_{\rm c}(\varepsilon)$ 
with $\rho_{\rm f}(\varepsilon)\equiv\sum_{i=1}^{24}A_{ii}^{\rm ff}(\varepsilon)$ 
and $\rho_{\rm c}(\varepsilon)\equiv\sum_{\xi=1}^{48}A_{\xi\xi}^{\rm cc}(\varepsilon)$,  
where spectral function $A_{bb}^{\rm aa}(\varepsilon)$ is defined as 
$A_{bb}^{\rm aa}(\varepsilon)\equiv-\frac{1}{\pi{N_{\rm L}}}\sum_{\bf k}{\rm Im}G_{bb}^{\rm aa \ R}({\bf k}, \varepsilon)$ 
with 
the retarded Green function for quasiparticles 
$G^{\rm R}({\bf k}, \varepsilon)\equiv(\varepsilon+i\delta-\tilde{H}_{\bf k})^{-1}$ and $\delta=0.01$. 
Here, $\tilde{H}_{\bf k}$ is given by $\tilde{H}=\sum_{\bf k}\tilde{H}_{\bf k}$. 
Since $\varepsilon_{\rm f}$ is located at rather deep position from the Fermi level, the present state is considered to 
correspond to pressurized approximant crystal. 
Actually, 
at the Fermi level, the f-component is dominant with 
$\rho_{\rm f}(\varepsilon_{\rm F})$ being sharing $79.5\%$ of $\rho_{\rm tot}(\varepsilon_{\rm F})$. 
The conduction bands of 3p holes on the 4th, 2nd, and 5th shells hybridize with 4f holes 
with each DOS at $\varepsilon_{\rm F}$ sharing $15.7\%$, $2.8\%$, and $2.0\%$ of 
$\rho_{\rm tot}(\varepsilon_{\rm F})$, 
respectively, forming three hybridized bands at $\varepsilon_{\rm F}$ 
below pseudo-gap-like DOS around $\varepsilon\sim 0.3$ in Fig.~\ref{fig:band_DOS}(b).  
Almost flat dispersions around $\varepsilon\approx-1.2$ and $1.0$ 
in Fig.~\ref{fig:band_DOS}(a) are reflected in 
marked DOS's in Fig.~\ref{fig:band_DOS}(b), respectively, which are mainly due to 
splitting of 3p bands on the 2nd shell [see Fig.~\ref{fig:band_DOS}(c)]. 

\begin{figure}
\includegraphics[width=7.5cm]{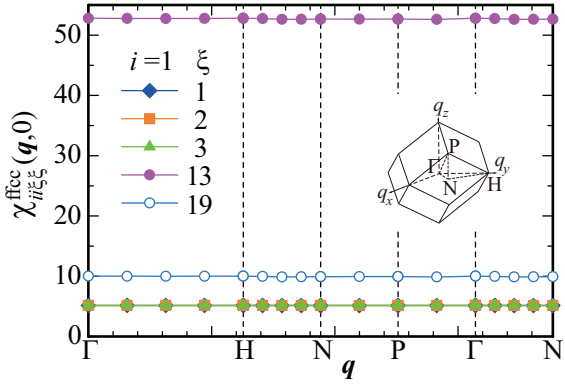}
\caption{(color online) 
$\chi^{\rm ffcc}_{ii\xi\xi,\sigma}({\bf q},0)$ 
along high-symmetry lines 
for $i=1$ and $\xi=1$ (filled diamond), $2$ (filled square), $3$ (filled triangle), 
$13$ (filled circle), and $19$ (open circle) at $T=0.0001$ calculated in $N_{\rm L}=8^3$ 
and $n_{\rm M}=2^{15}$ for $t_2=1.0$, $t_{5}'=0.2$, $V_0=0.13$, $\varepsilon_{\rm f}=-0.4$, 
and $U=\infty$ at $\bar{n}=1$. 
Note that data for $\xi=1$ and 2 are degenerated. 
Inset shows bcc Brillouin zone. 
}
\label{fig:chi_ffcc}
\end{figure}

The dynamical susceptibility for the charge transfer between 4f and conduction holes 
\begin{eqnarray}
\chi^{\rm ffcc}_{ii\xi\xi,\sigma}({\bf q},i\omega_{m})=
-\frac{T}{N_{\rm L}}\sum_{n{\bf k}}G_{ii,\sigma}^{\rm ff}({\bf k}+{\bf q},i\varepsilon_{n}+i\omega_{m})
G_{\xi\xi,\sigma}^{\rm cc}({\bf k},i\varepsilon_{n}) 
\label{eq:chi_ffcc}
\end{eqnarray}
is calculated by using $G({\bf k},i\varepsilon_{n})\equiv(i\varepsilon_{n}-\tilde{H}_{\bf k}+\mu)^{-1}$ 
with 
$\varepsilon_{n}=(2n+1)\pi{T}$ and the chemical potential $\mu$, and $\omega_{m}=2m\pi{T}$. 
The $\bf q$ dependence of $\chi^{\rm ffcc}_{ii\xi\xi,\sigma}({\bf q},0)$ calculated at $T=0.0001$ with 
the number of Matsubara frequency $n_{\rm M}=2^{15}$ being kept is shown in Fig.~\ref{fig:chi_ffcc}. 
A remarkable result is that almost flat-momentum dependence appears 
in the charge-transfer mode between 
the $i=1$st Yb site [see Fig.~\ref{fig:Yb_cluster}(c)] and 
the $\xi=1$st, $2$nd, $3$rd Al sites [see Fig.~\ref{fig:Yb_cluster}(b)], 
$\xi=13$th Al site [see Fig.~\ref{fig:Yb_cluster}(d)], and $\xi=19$th Al site [see Fig.~\ref{fig:Yb_cluster}(e)], which are  
connected via $H_{\rm hyb}$ in Eq.~(\ref{eq:Hhyb}).
Emergence of almost flat mode is considered to be ascribed to strong local correlation 
effect by $U=\infty$~\cite{WM2010,Miyake_2007,WM2012}.

The reason why the charge-transfer mode between the 3rd and 4th shells   
is extraordinary enhanced [see the $(i,\xi)=(1,13)$ data in Fig.~\ref{fig:chi_ffcc}]
is due to the strongest f-c hybridization, $|V_{3-4}|>|V_{2-3}|, |V_{3-5}|$, 
arising from the shortest Yb-Al distance and 
existence of the DOS around $\varepsilon_{\rm F}$ 
[see $A_{\xi\xi}^{\rm cc}(\varepsilon_{\rm F})$ for $\xi=13$ in Fig.~\ref{fig:band_DOS}(c)]. 
Maximum of $\chi_{ii\xi\xi}^{\rm ffcc}({\bf q},0)$ for $i=1$ and $\xi=13$ 
is located at the $\Gamma$ point, ${\bf q}={\bf 0}$. 

To clarify how the locality of the charge-transfer fluctuation 
affects the quantum criticality, 
and also to get insight into the mechanism of emergence of the $T/B$ scaling, 
let us focus on the charge-transfer mode between the N.N. Yb 
on the 3rd shell and Al on the 4th shell  
since it is overwhelmingly dominant in Fig.~\ref{fig:chi_ffcc} [see the $(i,\xi)=(1,13)$ data]. 
Then, we apply the recently-developed mode-coupling theory for critical valence fluctuations 
under magnetic field~\cite{WM2014} to the present system, 
which starts from the Hamiltonian 
\begin{eqnarray}
{\cal H}=H+H_{U_{\rm fc}}+H_{\rm Zeeman}, 
\label{eq:H_field}
\end{eqnarray}
where the charge-transfer fluctuation, i.e., Yb-valence fluctuation, 
is caused by the inter-orbital Coulomb repulsion~\cite{OM2000,Miyake_2007,WM2010} 
\begin{eqnarray}
H_{U_{\rm fc}}=U_{\rm fc}\sum_{\langle ji,j'\xi\rangle\sigma\sigma'}
n_{ji\sigma}^{\rm f}n_{j'\xi\sigma'}^{\rm c}. 
\label{eq:Ufc}
\end{eqnarray}
Here, $\langle ji,j'\xi\rangle$ denotes the N.N. pair between Yb  
on the 3rd shell and Al on the 4th shell. 
The Zeeman term is given by 
\begin{eqnarray}
H_{\rm Zeeman}=
-h\sum_{j=1}^{N_{\rm L}}
\left[\sum_{i=1}^{24}S_{ji}^{{\rm f}z}
+\sum_{\xi=1}^{48}S_{j\xi}^{{\rm c}z}
\right],  
\label{eq:H_Zeeman}
\end{eqnarray}
where $h$ is magnetic field, and $S_{ji}^{{\rm f}z}\equiv\frac{1}{2}(n_{ji\uparrow}^{\rm f}-n_{ji\downarrow}^{\rm f})$ 
and $S_{j\xi}^{{\rm c}z}\equiv\frac{1}{2}(n_{j\xi\uparrow}^{\rm c}-n_{j\xi\downarrow}^{\rm c})$ .

Taking into account the mode-coupling effects 
of the charge-transfer fluctuation 
up to the 4th order of $U_{\rm fc}$ 
in the action $S[\varphi]$ for ${\cal H}$, which is derived by introducing the 
Stratonovich-Hubbard transformation for $H_{U_{\rm fc}}$ (see Ref.~\citen{WM2014} for detail), 
we construct the action for the Gaussian fixed point. 
By using Feynman's inequality~\cite{Feynman}, 
the free energy for $\cal H$ is evaluated as 
$F\le F_{\rm eff}+T\langle S-S_{\rm eff}\rangle_{\rm eff}\equiv\tilde{F}(\eta)$, 
where 
$S_{\rm eff}$ is the effective action for the best Gaussian, 
$S_{\rm eff}[\varphi]=\frac{1}{2}\sum_{\sigma}\sum_{{\bf q},m}
\chi_{{\rm v}\sigma}({\bf q},{i}\omega_{m})^{-1}
|\varphi_{\sigma}({\bf q},{i}\omega_{m})|^2$. 
Here, the valence susceptibility 
$\chi_{{\rm v}\sigma}({\bf q},{i}\omega_{m})$ 
is defined as 
\begin{eqnarray}
\chi_{{\rm v}\sigma}({\bf q},{i}\omega_{m})^{-1}
\approx\eta+A_{\sigma}q^2+C_{\sigma}\frac{|\omega_{m}|}{q}. 
\label{eq:chi_v}
\end{eqnarray}
and $F_{\rm eff}$ is the free energy for the best Gaussian.
By optimizing $\tilde{F}(\eta)$ by $\eta$, $\frac{d\tilde{F}(\eta)}{d\eta}=0$, 
the mode-coupling equation under magnetic field 
is derived in the $A_{\sigma}q_{\rm B\sigma}^{2}\lsim\eta$ region as 
\begin{eqnarray}
& &\sum_{\sigma}A_{\sigma}q_{{\rm B}\sigma}^4\frac{T_{0\sigma}}{T_{{\rm A}\sigma}^2}
\left(
1+\frac{v_{4\sigma}q_{{\rm B}\sigma}^3}{\pi^2}\frac{T_{0\sigma}}{T_{{\rm A}\sigma}^2}
\right)
\nonumber
\\
&{\times}&
\left[
y_{0\sigma}-\tilde{y}_{\sigma}+\frac{3}{2}y_{1\sigma}t_{\sigma}
\left\{
\frac{x_{\rm c}^3}{6\tilde{y}_{\sigma}}
-\frac{1}{2\tilde{y}_{\sigma}}
\int_{0}^{x_{\rm c}}\frac{x^3}{x+\frac{t_{\sigma}}{6\tilde{y}_{\sigma}}}dx
\right\}
\right]
\nonumber
\\
&{\times}&
\left[
C_{2\sigma}+\frac{t_{\sigma}x_{\rm c}^3}{3\tilde{y}_{\sigma}^2}
-\frac{t_{\sigma}}{\tilde{y}_{\sigma}^2}
\int_{0}^{x_{\rm c}}\frac{x^4}{\left(
x+\frac{t_{\sigma}}{6\tilde{y}_{\sigma}}
\right)^2
}dx
\right]
=0, 
\label{eq:VSCReq}
\end{eqnarray}
where $\tilde{y}_{\sigma}=y\frac{A}{A_{\sigma}}\left(\frac{q_{\rm B}}{q_{{\rm B}\sigma}}\right)^2$, 
$t_{\sigma}=\frac{T}{T_{0\sigma}}$, 
$T_{0\sigma}=\frac{A_{\sigma}q_{{\rm B}\sigma}^3}{2{\pi}C_{\sigma}}$, 
and  
$T_{{\rm A}\sigma}=\frac{Aq_{{\rm B}\sigma}^2}{2}$ 
with $q_{{\rm B}\sigma}$ being the Brillouin zone.  
Here, $y$ is defined as $y\equiv\frac{\eta}{Aq_{\rm B}^2}$, and the 
dimensionless variable and its cutoff are defined as    
$x\equiv q/q_{\rm B}$ and $x_{\rm c}\equiv q_{\rm c}/q_{\rm B}$, respectively.  
Note that $A$, $C$, and $q_{\rm B}$ are the zero-field values of $A_{\sigma}$, 
$C_{\sigma}$, and $q_{{\rm B}\sigma}$, respectively. 
The parameters $y_{0\sigma}$ and $y_{1\sigma}$ are given by
\begin{eqnarray}
y_{0\sigma}&=&\frac{\frac{\eta_{0\sigma}}{A_{\sigma}q_{{\rm B}\sigma}^2}
+v_{4\sigma}\frac{T_{0\sigma}}{T_{{\rm A}\sigma}^2}\frac{q_{{\rm B}\sigma}^3}{\pi^2}C_{1\sigma}}
{1+v_{4\sigma}\frac{T_{0\sigma}}{T_{{\rm A}\sigma}^2}\frac{q_{{\rm B}\sigma}^3}{\pi^2}C_{2\sigma}}, 
\label{eq:y0}
\\
y_{1\sigma}&=&
\frac{v_{4\sigma}\frac{T_{0\sigma}}{T_{{\rm A}\sigma}^2}\frac{4q_{{\rm B}\sigma}^3}{3\pi^2}}
{1+v_{4\sigma}\frac{T_{0\sigma}}{T_{{\rm A}\sigma}^2}\frac{q_{{\rm B}\sigma}^3}{\pi^2}C_{2\sigma}}, 
\label{eq:y1}
\end{eqnarray}
respectively, where $\eta_{0\sigma}$ is defined as 
$\eta_{0\sigma}\equiv U_{\rm fc}\left[1-U_{\rm fc}\chi_{ii\xi\xi}^{\rm ffcc}({\bf 0},0)\right]$ 
and 
the mode-coupling constant of the 4th order $v_{4\sigma}$ is calculated as   
\begin{eqnarray}
v_{4\sigma}=\frac{U_{\rm fc}^4}{4}
\left[
\frac{T}{2N_{\rm L}}\sum_{n}\sum_{\bf k}
G_{\xi\xi,\sigma}^{\rm cc}({\bf k},{i}\varepsilon_{n})^2
G_{ii, \sigma}^{\rm ff}({\bf k},{i}\varepsilon_{n})^2
\right. 
\ \ \ \ \ \ \ \ \ \ \ \ \ \ \ \ \ \ \ 
\nonumber
\\
+\left.\frac{T}{N_{\rm L}}\sum_{n}\sum_{\bf k}
G_{\xi{i},\sigma}^{\rm cf}({\bf k},{i}\varepsilon_{n})
G_{ii,\sigma}^{\rm ff}({\bf k},{i}\varepsilon_{n})
G_{i\xi,\sigma}^{\rm fc}({\bf k},{i}\varepsilon_{n})
G_{\xi\xi,\sigma}^{\rm cc}({\bf k},{i}\varepsilon_{n})
\right].  
\label{eq:v4}
\end{eqnarray}
The constants 
$C_{1\sigma}$ and $C_{2\sigma}$ are given by   
$C_{1\sigma}=\int_{0}^{x_{\rm c}}dxx^3\ln\left|
\frac{(A_{\sigma}q_{{\rm B}\sigma}^2x^3)^2+(C_{\sigma}\omega_{\rm c}/q_{{\rm B}\sigma})^2}{(A_{\sigma}q_{{\rm B}\sigma}^2x^3)^2}
\right|
$ 
and 
$C_{2\sigma}=2(C_{\sigma}\omega_{\rm c})^2
\int_{0}^{x_{\rm c}}dx\frac{x}{(A_{\sigma}q_{{\rm B}\sigma}^3x^3)^2+(C_{\sigma}\omega_{\rm c})^2}
$, respectively. 

To proceed the calculation in Fig.~\ref{fig:chi_ffcc} to the mode-coupling theory for 
the approximant crystal, 
here we try to estimate the momentum and frequency dependence 
of $\chi_{ii\xi\xi}^{\rm ffcc}({\bf q},i\omega_{m})$ 
in the vicinity of 
$\chi_{ii\xi\xi}^{\rm ffcc}({\bf 0},0)$, as follows: 

The ${\bf q}^2$-coefficient is evaluated as 
\begin{eqnarray}
\chi_{ii\xi\xi,\sigma}^{\rm ffcc}({\bf q}_{\nu},0)=\chi_{ii\xi\xi,\sigma}^{\rm ffcc}({\bf 0},0)
-A_{\nu,\sigma}{\bf q}_{\nu}^2,
\label{eq:Aav}
\end{eqnarray}
where ${\bf q}_{\nu}=\frac{2\pi}{a}(\frac{2}{N_1},0,0)$, $\frac{2\pi}{a}(0,\frac{2}{N_2},0)$, $\frac{2\pi}{a}(0,0,\frac{2}{N_3})$, $\frac{2\pi}{a}(\frac{1}{N_1},\frac{1}{N_2},0)$, $\frac{2\pi}{a}(0,\frac{1}{N_2},\frac{1}{N_3})$, $\frac{2\pi}{a}(\frac{1}{N_1},0,\frac{1}{N_3})$ 
with $a=14.5$~\AA \ being a lattice constant in the $N_{\rm L}=N_1N_2N_3$ system. 
Then, we evaluate $A_{\sigma}$ in Eq.~(\ref{eq:chi_v}) 
as $A_{\sigma}\approx U_{\rm fc}^2A_{{\rm av},\sigma}$ 
by employing the averaged value  
$A_{{\rm av},\sigma}=\frac{1}{6}\sum_{\nu=1}^{6}A_{\nu,\sigma}$. 

As for the frequency dependence, 
\begin{eqnarray}
\chi_{ii\xi\xi,\sigma}^{\rm ffcc}({\bf q}_{\nu},i\omega_1)-\chi_{ii\xi\xi,\sigma}^{\rm ffcc}({\bf q}_{\nu},0)
\approx-C_{\nu,\sigma}\frac{|\omega_1|}{q_{\nu}},
\label{eq:Cav}
\end{eqnarray}
where $q_{\nu}=|{\bf q}_{\nu}|$ and here inter-band contribution is neglected since 
intra-band contribution is dominant. 
Then, we evaluate $C_{\sigma}$ in Eq.~(\ref{eq:chi_v}) as 
$C_{\sigma}\approx U_{\rm fc}^{2}C_{{\rm av},\sigma}$ by employing the averaged value  
$C_{{\rm av},\sigma}=\frac{1}{6}\sum_{\nu=1}^{6}C_{\nu,\sigma}$. 

In this way, the mode-coupling theory can be applied to the approximant crystal.
The procedure of the calculation is as follows:  
First, we solve the slave-boson mean-field equations [Eqs.~(\ref{eq:MF1}) and (\ref{eq:MF2})]  
at $T=0$ for a given set of parameters: $t_2$, ${t_5}'$, $V_0$, $\varepsilon_{\rm f}$, 
$U=\infty$, and $h$ at filling $\bar{n}$. 
Second, we calculate $\chi_{ii\xi\xi,\sigma}^{\rm ffcc}({\bf q},i\omega_m)$ 
in Eq.~(\ref{eq:chi_ffcc}) and the $\left[...\right]$ part in Eq.~(\ref{eq:v4}) 
by using the mean-field solution.   
Then, we obtain $\eta_{0\sigma}$ and $v_{4\sigma}$ for a given $U_{\rm fc}$. 
Third, by using $y_{0\sigma}$ and $y_{1\sigma}$ obtained from Eqs.~(\ref{eq:y0}) 
and (\ref{eq:y1}), respectively,   
we solve the mode-coupling equation [Eq.~(\ref{eq:VSCReq})] for critical valence fluctuations 
and finally obtain the solution $y(T)$. 

At the QCP of valence transition, 
the magnetic susceptibility $\chi$ as well as the valence susceptibility $\chi_{\rm v}({\bf 0},0)$ 
diverges with the same singularity, 
$\chi\propto\chi_{\rm v}({\bf 0},0)\propto y^{-1}$, 
since the many body effect for $U_{\rm fc}$ common to both $\chi$ and $\chi_{\rm v}({\bf 0},0)$ is enhanced 
near the QCP~\cite{WM2010}. 

By setting the parameters used in Fig.~\ref{fig:chi_ffcc}, we perform the above procedure. 
Here we calculate $\chi^{\rm ffcc}_{ii\xi\xi}({\bf q},i\omega_m)$ and $v_{4\sigma}$ at $T=0.0001$ 
with $\mu$ being determined so as to satisfy $\bar{n}=1$ and temperature evolution of $y(T)$ 
is obtained by solving Eq.~(\ref{eq:VSCReq}). 
Since approximant crystal has large unit cell with $a=14.5$~\AA, $q_{\rm B}=\frac{2\pi}{a}$ 
is one-order of magnitude smaller than those of usual periodic crystals.  
This makes characteristic temperature of critical valence fluctuation 
$T_{0}\equiv\frac{Aq_{\rm B}^3}{2\pi C}$ reduced. 
Indeed we obtained $T_{0}=1.4\times10^{-4}$, which is four-order of magnitude 
smaller than the band width (see Fig.~\ref{fig:band_DOS}).
The QCP of valence transition is identified as 
$U_{\rm fc}=0.0192$ where $y(0)$ becomes zero. 

\begin{figure}
\includegraphics[width=7.5cm]{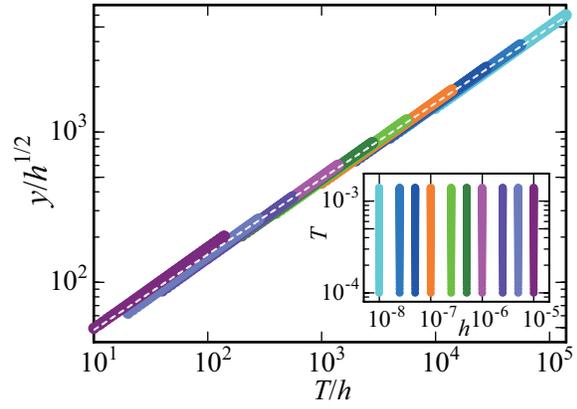}
\caption{(color online) 
Scaling of the data for $0.0001\le{T}\le0.0014$ and $10^{-8}\le{h}\le10^{-5}$. 
Inset shows the $T$-$h$ range where the scaling applies. 
The dashed line represents the fitting function $c(T/h)^\zeta$ with $\zeta=0.503$. 
The data were obtained 
for $t_2=1.0$, $t_{5}'=0.2$, $V_0=0.13$, $\varepsilon_{\rm f}=-0.4$, $U_{\rm fc}=0.0192$, 
and $U=\infty$ at $\bar{n}=1$ in $N_{\rm L}=8^3$ and $n_{\rm M}=2^{15}$. 
}
\label{fig:scaling}
\end{figure}

Figure~\ref{fig:scaling} shows the result of the solution of Eq.~(\ref{eq:VSCReq}) 
for $0.0001\le T\le 0.0014$ and $10^{-8}\le h \le 10^{-5}$ at the QCP. 
We find that the data seem to  
fall down to a single scaling function of $T/h$ over four decades. 
\begin{eqnarray}
y=h^{1/2}\phi\left(\frac{T}{h}\right). 
\label{eq:scaling}
\end{eqnarray}
The least-square fit for the large $T/h$ regime of $10^{2}\le T/h \le 1.4\times 10^5$ 
gives $\frac{y}{h}=c\left(\frac{T}{h}\right)^\zeta$ with $\zeta=0.503$, which is shown 
as a dashed line. 
This indicates that scaling function for $x\gg 1$ has the form of $\phi(x)=cx^{1/2}$. 
In this $T/h\gg1$ region, we have $y/h^{1/2}\approx c(T/h)^{1/2}$, i.e., $y\approx T^{1/2}$. 
This is caused by critical valence fluctuation with strong locality, giving rise to 
the non-Fermi-liquid regime. 
This result indicates that the magnetic susceptibility behaves as $\chi\approx T^{-1/2}$ for $h\to 0$, which 
has been observed in quasicrystal Yb$_{15}$Al$_{34}$Au$_{51}$~\cite{Deguchi}. 
Our result implies that 
even in the approximant crystal, the same behavior is expected to appear 
when pressure is applied. 

As $T/h$ decreases, the data tend to show deviation from the dashed line as seen 
in Fig.~\ref{fig:scaling}. 
This reflects suppression of valence susceptibility 
by magnetic field and indicates crossover to 
the Fermi liquid regime toward $T/h\ll 1$. 

Emergence of the $T/h$ scaling behavior is ascribed to presence of small characteristic 
temperature of critical valence fluctuation $T_0$.  
In case that $T_0$ is below (or at least comparable to) the lowest temperature, 
$t/y\gg 1$ holds where all the terms with $y$ and $t$ in Eq.~(\ref{eq:VSCReq}) can be expressed as 
scaling forms of $y/h^{1/2}$ and $t/h$, respectively~\cite{WM2014}. 
We have confirmed that this is the case for all the data used in Fig.~\ref{fig:scaling}. 
Hence, realization of small $T_0$ coming from almost flat-$\bf q$ charge-transfer mode 
(see Fig.~\ref{fig:chi_ffcc}) assisted by small Brillouin zone $q_{\rm B}$ 
reflecting large unit cell is the main reason for emergence of the $T/h$ scaling 
as well as $\chi\sim T^{-1/2}$ for the zero-field limit. 

In reality, there exist $3p_x$, $3p_y$, $3p_z$ bands from Al and $6s$ band from Au.  
Those conduction bands are folded into small Brillouin zone and hybridizations between them 
each other give rise to many splits into bonding and antibonding bands~\cite{Fujiwara}. 
Hence, conduction bands themselves have the flat-$\bf q$ nature. 
Hybridization between f and their bands is expected to further promote locality of 
valence fluctuations. 

On the basis of the solution obtained in Eq.~(\ref{eq:VSCReq}),  
it is shown that  
critical Yb-valence fluctuation causes 
a new type of criticality such as   
NMR relaxation rate $(T_{1}T)^{-1}\propto\chi\sim T^{-0.5}$, 
specific heat $C_{\rm e}/T\sim-\log{T}$, and 
$T$-linear resistivity for $T/T_{0}\gsim 1$~\cite{WM2010,WM2014}. 
Applying pressure to the approximant crystal is considered to make the f-hole level decrease and $U_{\rm fc}$ increase. Hence, the quantum valence criticality is expected to appear under pressure. 
The quasicrystal corresponds to the infinite limit of the unit-cell size in the approximant crystal. Then, the quasicrystal is regarded as the system with the small limit of the Brillouin zone, $q_{\rm B}\to 0$, giving rise to vanishing characteristic temperature of critical Yb-valence fluctuation, i.e., $T_{0}\equiv\frac{Aq_{\rm B}^3}{2\pi C}\to 0$. Hence, $T_{0}$ is expected to be smaller than the measured lowest temperature, so that present mechanism is considered to capture the essence of the origin of the unconventional critical phenomena as above including the $T/B$ scaling in $\chi$ observed in Yb$_{15}$Al$_{34}$Au$_{51}$.

{\bf Acknowledgements} 
We thank N. K. Sato, K.~Deguchi, S. Matsukawa, T.~Ishimasa, and T.~Watanuki 
for showing us experimental data with enlightening discussions on their analyses. 
This work was supported by Grants-in-Aid for Scientific Research (No. 24540378, No. 25400369 and No. 16H01077A01) from the Japan Society for the Promotion of Science (JSPS).

\end{document}